\documentclass[twocolumn]{aastex63}
\shortauthors{Villarroel et al.}
\graphicspath{{./}{figures/}}

\begin{document}

\title{Is there a background population of high-albedo objects in geosynchronous orbits around Earth?}

\author{Beatriz Villarroel}
\affiliation{Nordita, KTH Royal Institute of Technology and Stockholm University, Hannes Alfv\'ens v\"ag 12, SE-106 91 Stockholm, Sweden}
\affiliation{Instituto de Astrof\'isica de Canarias, Avda V\'ia L\'actea S/N, La Laguna, E-38205, Tenerife, Spain}
\author{Enrique Solano}
\affiliation{Departamento de Astrof\'isica, Centro de Astrobiolog\'ia (CSIC/INTA), PO Box 78, E-28691, Villanueva de la Ca\~{n}ada, Spain}
\affiliation{Spanish Virtual Observatory}
\author{Hichem Guergouri}
\affiliation{Science of the Matter Division. Research Unit in Scientific Mediation, CERIST, Constantine, Algeria}
\author{Alina Streblyanska}
\affiliation{Instituto de Astrof\'isica de Canarias, Avda V\'ia L\'actea S/N, La Laguna, E-38205, Tenerife, Spain}
\author{Lars Mattsson}
\affiliation{Nordita, KTH Royal Institute of Technology and Stockholm University, Hannes Alfv\'ens v\"ag 12, SE-106 91 Stockholm, Sweden}
\author{Rudolf E. B\"ar}
\affiliation{Institute for Particle Physics and Astrophysics, ETH Zurich, Wolfgang-Pauli-Strasse 27, CH-8093 Zurich, Switzerland}
\author{Jamal Mimouni}
\affiliation{Depart. of Physics, Univ. of Constantine-1, LPMPS \& CERIST, Constantine, Algeria}
\author{Stefan Geier}
\affiliation{Gran Telescopio Canarias (GRANTECAN), Cuesta de San Jos\'{e} s/n, 38712 Bre\~{n}a Baja, La Palma, Spain}
\author{Alok C. Gupta}
\affiliation{Aryabhatta Research Institute of Observational Sciences (ARIES), Manora Peak, Nainital, 263 001, India}
\author{Vanessa Okororie}
\affiliation{Center for Basic Space Science, National Space Research and Development Agency, Enugu-Nigeria.
}
\author{Khaoula Laggoune}
\affiliation{Sirius Astronomy Association, Algeria
}
\author{Matthew E. Shultz}
\affiliation{Department of Physics and Astronomy, University of Delaware, USA}
\author{Robert A. Freitas Jr.}
\affiliation{Institute for Molecular Manufacturing, Palo Alto CA, United States}
\author{Martin J. Ward}
\affiliation{Centre for Extragalactic Astronomy, Department of Physics, Durham University, South Road, Durham, DH1 3LE, UK}

\begin{abstract}
Old, digitized astronomical images taken before the human spacefaring age offer a unique view of the sky devoid of known artificial satellites. In this paper, we have carried out the first optical searches ever for non-terrestrial artifacts near the Earth following the method proposed in \cite{Villarroel2022a}. We use images contained in the First Palomar Sky Survey to search for simultaneous (during a plate exposure time) transients that in addition to being point-like, are aligned. 
We provide a shortlist of the most promising candidates of aligned transients, that must be examined with the help of a microscope to separate celestial sources from plate defects with coincidentally star-like brightness profiles. We further explore one possible, but not unique, interpretation in terms of fast reflections off high-albedo objects in geosynchronous orbits around Earth. If a future study rules out each multiple transient candidate, the estimated surface density becomes an upper limit of $<10^{-9}$ objects km$^{-2}$ non-terrestrial artifacts in geosynchronous orbits around Earth. Finally, we conclude that observations and analysis of multiple, simultaneously appearing and vanishing light sources on the sky merit serious further attention, regardless of their origin.

\end{abstract}



\section{Introduction}
\label{sec:intro}
Digitized sky surveys have broadened the time window in which we can study the changes of our sky. Programs such as the Digital Access to a Sky Century at Harvard \citep[DASCH;][]{Grindlay2012}, the Digital Sky Survey\footnote{https://archive.stsci.edu/cgi-bin/dss\_form/} (DSS), Ukraine Virtual Observatory \citep[JDA UkrVO;][]{Vavilova2012,Vavilova2017} and \textit{Carte du Ciel} go back not only a few decades in their data collection, but contain images of the sky as old as 150 years.
While photographic plates are no longer used for the large astronomical surveys as they were replaced by the significantly faster and more responsive CCDs, the old archive images can serve several purposes. For example, they can be used to trace the variability of a particular object over a time period of several decades or even a hundred years, assuming that the object is bright enough to be detected. 

Another use of these catalogues is for the search for vanishing stars. In the Vanishing \& Appearing Sources during a Century of Observations \citep[VASCO;][]{Villarroel2016,Villarroel2020} project we use images of the sky taken in the 1950s and compare these with modern surveys to look for possible sources that may have vanished. In order to search for vanishing sources, the VASCO project employs two complementary approaches: first, an automated procedure \citep{Solano2021} that directly searches the many terabytes of digitized image data from the First and Second Palomar Sky Surveys (POSS-I and POSS-II) for transients, and second, a citizen science project \citep{VillarroelCSP} that allows the users to indicate any object they find to be interesting. The Spanish Virtual Observatory\footnote{http://svo.cab.inta-csic.es} with its many appropriate software tools has greatly facilitated these searches.

Recently, a very intriguing finding emerged from the VASCO project \citep{Villarroel2021a}: nine star-like objects that appeared and vanished simultaneously were found on a 1950's epoch  photographic plate in POSS-I. These nine transients could not be detected half an hour earlier on another plate, or six days later. All known astrophysical explanations were considered but found not to be plausible. The surface density of the simultaneously appearing and vanishing objects was far too high to be assigned to any established phenomenon. It could not be determined whether such a phenomenon could be some type of false positive with coincidentally star-like appearances, caused by some type of contamination of the photographic plate e.g. caused by radioactive particles from atomic bomb tests \citep[][]{Webb1949}, or if it was a genuine observation. For the latter case, a simple consistent explanation is that it is caused by solar reflections off flat and highly reflective objects in Geosynchronous Earth Orbits (GEOs) around the Earth. If correct, this would be of utmost significance in searches for non-terrestrial artifacts \citep[NTAs;][]{Bracewell,Freitas,Valdes,Kopparapu}. But even if real, it might be an entirely different and unknown phenomenon to us, causing a cluster of light sources to appear and simultaneously vanish on the sky. If so, the light sources do not need to be aligned.

In a related paper \cite{Villarroel2022a} proposed a methodology for how to search for other, unambiguous signatures of solar reflections from objects in GEOs by using photographic plates exposed before the satellite era commenced in 1957. Such signatures include several transients along a line. Natural transients are detected at a several of orders magnitude lower rate, and even finding two single natural transients a few arcminutes from each other in space at the same time is extremely unlikely. But transient events caused by solar reflections off flat objects with high albedo at high altitude around the Earth, could create an effect of multiple, simultaneously occurring transients in a long-exposure image. The detection of multiple transients in an image is common in today's surveys. In fact more than 80\% of all transients detected by the automated transient surveys are caused by this phenomenon due to the presence of space debris and satellites, see e.g. \cite{Nir2020,McDowell}, some of which are multiple transients in a single image. The typical apparent magnitudes of these glints are $\sim$ 8-10 magnitudes. The rate of these glints caused by human artifact space debris is so high when observing the sky near to the equator, around 1800 glints h$^{-1}$ sky$^{-1}$ \citep{McDowell}, that other phenomena with the same observational signature would be swamped  in modern surveys unless explicitly searched for.

The proposed method described in \cite{Villarroel2022a} searches for events appearing as ``simultaneous transients'' in long-exposure photographic plates that in addition to appearing in the same image are also aligned along a straight line. This criterion helps to distinguish between multiple transient events caused by high albedo objects from other causes of random point sources on an image (celestial or instrumental). In an image with nine simultaneous transients inside a $10 \times 10$ arcmin$^2$ box, the probability that several point sources will align by chance is very low, corresponding to statistical significance of from 2.5$\sigma$ (4-point alignments) to 3.9$\sigma$ (5-points). With fewer transients inside the box, the probability of finding a chance alignment strongly decreases, which means that we might also consider 3-point alignments if the total number of transients $N$ is small. Candidates with the lowest probability of being chance alignments should be considered the most promising candidates of multiple transients in the POSS-I dataset, rather than regarded as any conclusive proof of GEO glints. We suggest to examine these candidates carefully under a microscope, using the original photographic plates.

In this paper we carry out the test proposed by \cite{Villarroel2022a} using the automated list of POSS-I transients taken from \cite{Solano2021}. We find several candidates that we inspect in detail in Section \ref{sec:shortlist}. We use the existing observations of candidate alignments to deduce information about the basic geometry of the reflective objects, assuming they are real. We also used these examples to estimate their background density at the GEOs. Finally, we use this information to make predictions on how these reflective objects, \textit{assuming they are real}, can be found in ongoing and upcoming modern sky surveys.

\newpage
\section{Methods \& selection}\label{sec:methods}
We use the list of VASCO transients from \cite{Solano2021}, found on red POSS-I plates. The red POSS-I plates have average exposure times ranging from $45- 50$ minutes. We investigate grouped transients where several fall within the same box with each side ranging from a few arcminutes to $20-30$ arcmin in size (see typical sizes in Table \ref{shortlist}). For those cases we then investigate the correlation coefficient between the right ascensions and declinations of the transients, to determine if within positional errors they fall on a straight line.
We do this by only considering objects with a linear Pearson correlation coefficient $\alpha$ larger than abs($\alpha$) $>$ 0.99. Table \ref{Candidates} shows the number of cases with more than 3, 4, 5 or more transients in a row detected through this method. Each of these alignments can be found in image boxes of slightly different sizes due to the search method used.

\begin{table*}[ht]
\caption{{\bf Candidates.} We present the total number of candidates found in each region. In total 83 $r$-point alignments were found in the northern hemisphere, where $r$ is the number of aligned points. Note, that the sets $r \geq 4$ and $r \geq 5$ are subsets of $r \geq 3$. R.A. and Dec. in degrees.}
\centering
\begin{tabular}{c c c c c c}
\hline\hline
\multicolumn{3}{c}{Samples $\alpha >$ 0.99} \\
\hline\hline
Region & $r \geq 3$ & $r \geq 4$ & $r \geq 5$ & $r \geq 6$ \\[0.1ex]
0 $<$ R.A. $<$ 100, 0 $<$ Dec. $<$ 90 & 22 & 5 & - & - &\\[0.1ex]
100 $<$ R.A. $<$ 200, 0 $<$ Dec. $<$ 90 & 18 & 7 & - & - &\\[0.1ex]
200 $<$ R.A. $<$ 300, 0 $<$ Dec. $<$ 90 & 32 & 6 & 1 & - &\\[0.1ex]
300 $<$ R.A. $<$ 360, 0 $<$ Dec. $<$ 90 & 11 & 2 & 1 & - &\\[0.1ex]
\hline
Total & 83 & 20 & 2 & 0\\[0.1ex]
\hline
\end{tabular}
\label{Candidates}
\end{table*}

Images of all 83 candidates are given in the Appendix. Even by simple visual examination of the candidates, it is clear that duplets and triplets of transients are rather common among the examples.

Rather than examining each single $N \geq 3$ point alignment and determining whether this alignment is significant or not, our goal is to examine the most interesting cases at higher resolution. Often, but not always, the same plate has been scanned separately by DSS and SuperCosmos \citep{Hambly2001}. The latter has a higher digital resolution which improves examination of the shape of the object.

We have therefore downloaded new fits files from DSS and SuperCosmos of all candidates with $N \geq 4$. For each candidate we select an image box that includes the entire alignment. Obviously the larger the image box, the greater the chance of finding more ``point sources'' (real or plate defects). In several cases we discover that what was thought to be a point source in a DSS image, was in fact a transient of dubious shape (possibly a round plate flaw). Some other transients are apparently a result of scanning defects, and are not present in the corresponding SuperCosmos scan of the same plate.

Principally we consider cases that from the first visual inspection displayed at least four transients following a line. Next we investigate them using the SuperCosmos catalog. Of the candidates in the northern hemisphere, we focus on the five candidates in the final shortlist, listed in Table \ref{shortlist}.

\section{The shortlist}\label{sec:shortlist}

The shortlist in Table \ref{shortlist} shows the candidates. Each candidate is shown in Figures 1 - 5. Here we show only the transients themselves to assist the reader. The same images, but showing the actual alignments can be found in Figures 7 - 11. The alignments  differ in width, therefore a dashed double line is shown in some particular cases where the width of the stripe is larger than 1 arcsec.

In some cases, for example the objects marked with crosses in Candidate 3 and Candidate 5, it is not certain that every transient is a point source based on inspection of the images and slight asymmetries in the light profiles are present. Therefore, the alignment is a possible to be a combination of transients and plate defects. The reader can inspect the high resolution images from SuperCosmos\footnote{http://www-wfau.roe.ac.uk/sss/pixel.html}. We improve the astrometry for the images using the Terapix \textit{swarp} procedure. We measure the improved coordinates and the FWHM for each transient, see Table \ref{Measurements}.

In a few cases it is possible to derive more variants on the alignment, i.e. with either a 3-point or a 4-point alignment. In such cases we show both options separately in the image in the Figures 7 - 11. For the cases in the shortlist we estimate the probability of a chance alignment, see Section \ref{sec:statistics}.
\\
\section{Statistics}\label{sec:statistics}

For each of the interesting cases we consider the total number $N$ of transient-like objects found in the image field, the area $A$ of the image, the width of the strip $p_{max}$ needed to align the transients in the image, the length $d_{max}$ of the line and the number $r$ of objects that are aligned. 

As the box sizes are different for each case, we must estimate the expected number of alignments $\mu_{r}$ for the number of $r$-point alignments within a field $A$. As suggested in \cite{Villarroel2022a}, we use the generalization formula from \cite{Edmunds85},
\begin{equation}
    \label{eq:rpoint}
    \mu_r = \frac{\pi\,2^{r-2}\,n^r\,p_{\rm max}^{r-2}\,A} {\Gamma(r-1)}\int_0^{d_{\rm max}} x^{r-1}\,e^{-2x\,n\,p_{\rm max}}\,dx,
\end{equation}
where $\Gamma$ is the gamma function and $n = N/A$, $d_{\rm max}$ is the maximal length of the alignment (in arcmin), $p_{\rm max}$ is the width of the line (in arcmin) and $A$ is the area of the image in arcmin$^{2}$. As in \cite{Villarroel2022a} we use, for practical reasons, a limiting case of this generalization,
\begin{equation}
\label{eq:rpoint_approx}
    \mu_{r} \approx \frac{\pi\,2^{r-2}\,n^r\,p_{\rm max}^{r-2}\,d_{\rm max}^r\,A}{r\,(r-2)!}, \quad r = 3,4,5,...,
\end{equation}
which is good approximation when $2\,d_{\rm max}\,p_{\rm max}\,n \ll 1$ and simplifies the calculations considerably. For the present study equations (\ref{eq:rpoint}) and (\ref{eq:rpoint_approx}) should yield very similar results, since $2\,d_{\rm max}\,p_{\rm max}\,n \lesssim 0.01$ for all cases considered.

We apply equation (\ref{eq:rpoint_approx}) to calculate the expectation value $\mu_{r}$ for each case. We include all measurements in Table \ref{shortlist}. The short list includes both 3-point alignments and 4-point alignments. 
Since each candidate case only has one alignment, the 
probability is given by the expectation value, $p \sim \mu_{r}$. We can see that several of the cases are significantly statistically improbable ($3 - 4 \sigma$) to happen in a single image.

The probability estimate is also very sensitive to the total number of transients ($N$) present. This number depends strongly on the visual inspection that was made by blinking the POSS-I and POSS-II images in SAOImage DS9, taking into account the differences in depth. Any missed transients will change the value of $N$, and hence the estimated probability.

However, to estimate exactly the \textit{total} probability $p_{tot}$ of each single event to happen during our searches, two more factors influence the total probability. The first is the probability for obtaining a perfectly 1, 2, ... or $N\prime$ star-like plate flaws within the same area of an image. Given the rarity of encountering a star-like plate defect, and even less so with a matching FWHM as the normal stars of the same magnitude range in the field, it may be even more unusual to encounter 2, or 3, or 4 plate flaws that all have the same coincidental features and this lowers heavily the total probability of the event. The second factor is the total number of multiple transients in our dataset: if there are sufficiently many star-like plate flaws causing ``multiple transients'', some of these will line up. With an infinite dataset, any type of constellations will be found. This factor will, contrary to the first factor, increase the total probability for an event to occur.

Unfortunately, we have no grasp or means of estimating of either of the two factors. This means, a statistical proof for the existence of high-albedo objects at the GEO cannot be achieved based on our current information. In particular, we need to rule out defects by examination of the original plates.

It is easier to examine the effect of the choice of $p_{max}$ on the probability estimates for single images. The choice of $p_{max}$ depends on the science question of interest: are we interested whether the objects are truly aligned or whether they are just non-random? Showing non-randomness is all what is needed to argue for the authenticity of the points, but not necessarily enough to argue for that they truly are aligned as in the case of GEO glints. We use Table \ref{Measurements} to adopt other values of $p_{max}$, setting it equal to FWHM of the smallest star in an alignment (e.g. for Candidate 1, FWHM = 2.7 arcsec). Doing this, we see that all 3-point alignments are non-interesting events with $p > $ 0.05 (less significant than 2$\sigma$), with an exception of the border-line case of Candidate 2. This shows that for POSS-I data 
where the seeing in general is rather large, 3-point alignments of simultaneous transients do not provide significant proof against randomness. The only interesting cases remaining are the 4-point and 5-point cases, namely Candidates 3, 4 and 5, and yet both Candidate 3 and 5 show asymmetries in the shapes of the transients. And yet we could argue that they might not be as significant given the uncertainties previously described regarding the large data set we have examined to identify them.


\begin{table*}[ht]
\caption{{\bf Candidate shortlist.} We show the most interesting candidates emerging after 
the visual inspection. In some cases there could be different possibilities of $r$-point alignments, e.g. r = 3 or r = 4, and we show both possibilities marked by an asterisk (*). The given position coordinate corresponds to the transient marked with a cross (+) in each figure.}
\centering
\begin{tabular}{c c c c c c c c c cc}
\hline\hline
\multicolumn{10}{c}{Candidate shortlist} \\
\hline\hline
Candidate & Year & R.A. Dec. (sexag., J2000) & R.A. Dec. (deg, J2000) & $r$ & $N$ & $A$ [arcmin$^{2}$] & $p_{\rm max}$ [arcsec] & $d_{\rm max}$ [arcmin] & $\mu_{r} $ \\[3mm]
1 & 1954 & 02:29:33.71	+28:31:56.98 & 37.3904454 28.5324936 & 3 & 4 & 10 $\times$  10 & 1.0 & 5.8 & 0.044\\
2 & 1955 & 03:05:42.48 +07:58:29.60 & 46.4269814 7.9748892 & 3 & 5 & 10 $\times$  10 & 1.0 & 3.6 & 0.010\\
3 & 1954 & 03:08:27.13	+34:40:46.01 & 47.1130236 34.6794470 & 3 & 5 & 15 $\times$  15 - 16 & 2.0 & 9.9 & 0.194\\
" & " & "" & " & 5* & 5 & " & 15.0 & " & 0.002\\
4 & 1954 & 21:24:39.71	+68:31:30.04 & 321.1654740 68.5250111 & 3 & 6 & 12 $\times$  12 & 1.0 & 5.15 & 0.049 \\
" & - & " & " & 4* & 6 & " & 5.0 & " & 0.003 \\
5 & 1952 & 19:16:45.76	+51:28:52.40 & 289.1906854 51.4812217 & 3 & 5 & 10 $\times$  10 & 1.0 & 4.0 & 0.028\\
" & - & " & " & 5* & 5 & 10 $\times$  10 & 10.0 & 4.0 & 0.0001\\[3mm]
\hline
\end{tabular}
\label{shortlist}
\end{table*}

\begin{table*}[ht]
\caption{{\bf Measurements.} We list the astrometry-improved measurements for the objects inside the green circles
in Figures 1 - 5. Objects that are placed inside an alignment are marked with an asteriskm $\ast$.
The central objects presented in Table 2 are marked with a dagger ($\dagger$).
We show the FWHM in pixel and arcsec. The SuperCosmos resolution is 0.67 arcsec pixel$^{-1}$. The object
have an improved astrometry with help of Terapix swarp procedure, using zero-point calculations with SDSS as a reference field.} 
\centering
\begin{tabular}{c c c c c}
\hline\hline
\multicolumn{5}{c}{Candidates $1 - 5$} \\
\hline\hline
\hline
Object & R.A. & Dec. (sexag., J2000) & FWHM (pixel) & FWHM (arcsec) \\[3mm]
object1 & 2:29:37.57 & +28:36:31.58 & 4.0 & 2.7\\
object2$\ast$ & 2:29:21.38 & +28:36:57.89 & 7.2 & 4.8\\
object3$\ast$ & 2:29:21.76 & +28:36:49.09 & 7.6 & 5.1\\
object4$\dagger\ast$ & 2:29:33.80 & +28:31:56.83 & 4.1 & 2.7\\
\multicolumn{5}{c}{Date of observation= 1954-10-05} \\[3mm]
\hline
Object & R.A. & Dec. (sexag., J2000) & FWHM (pixel) & FWHM (arcsec) \\[3mm]
object1 & 3:05:52.34 & +8:00:16.97 & 3.8 & 2.5\\
object2$\dagger\ast$ & 3:05:42.46 & +7:58:30.22 & 10.0 & 5.7\\
object3$\ast$ & 3:05:42.81 & +7:58:20.56 & 5.9 & 4.0\\
object4$\ast$ & 3:05:50.24 & +7:55:33.86 & 4.4 & 2.9\\
\multicolumn{5}{c}{Date of observation= 1955-01-15} \\[3mm]
\hline
Object & R.A. & Dec. (sexag., J2000) & FWHM (pixel) & FWHM (arcsec) \\[3mm]
object1$\ast$ & 3:08:29.90 & +34:31:25.73 & 6.2 & 4.2\\
object2$\ast$ & 3:08:30.72 & +34:31:27.44 & 5.2 & 3.5\\
object3$\dagger\ast$ & 3:08:27.42 & +34:40:46.00 & 9.9 & 6.6\\
object4$\ast$ & 3:08:27.05 & +34:41:13.49 & 8.1 & 5.4\\
object5$\ast$ & 3:08:26.56 & +34:41:07.89 & 6.0 & 4.0\\
\multicolumn{5}{c}{Date of observation= 1954-12-22} \\[3mm]
\hline
Object & R.A. & Dec. (sexag., J2000) & FWHM (pixel) & FWHM (arcsec) \\[3mm]
object1 & 21:24:45.51 & +68:34:00.29 & 4.4 & 2.9\\
object2 & 21:24:44.59 & +68:34:01.20 & 4.6 & 3.1\\
object3$\ast$ & 21:24:47.62 & +68:31:58.92 & 4.4 & 2.9\\
object4$\dagger\ast$ & 21:24:39.72 & +68:31:31.22 & 8.9 & 6.0\\
object5$\ast$ & 21:24:38.18 & +68:31:27.97 & 5.0 & 3.4\\
object6$\ast$ & 21:24:03.94 & +68:29:14.36 & 4.6 & 3.1\\
\multicolumn{5}{c}{Date of observation= 1954-08-06} \\[3mm]
\hline
Object & R.A. & Dec. (sexag., J2000) & FWHM (pixel) & FWHM (arcsec) \\[3mm]
object1$\ast$ & 19:16:51.46 & +51:30:24.51 & 11.0 & 7.4  \\  
object2$\ast$ & 19:16:50.64 & +51:30:20.86 & 12.0 & 8.0\\
object3$\dagger\ast$ & 19:16:45.73 & +51:28:52.04 & 7.2 & 4.8\\
object4$\ast$ & 19:16:40.13 & +51:27:12.85 & 5.0 & 3.4\\
object5$\ast$ & 19:16:40.27 & +51:27:06.29 & 5.5 & 3.7\\
\multicolumn{5}{c}{Date of observation= 1952-07-28} \\[3mm]
\hline
\end{tabular}
\label{Measurements}
\end{table*}

\begin{figure*}
   \includegraphics[scale=0.2]{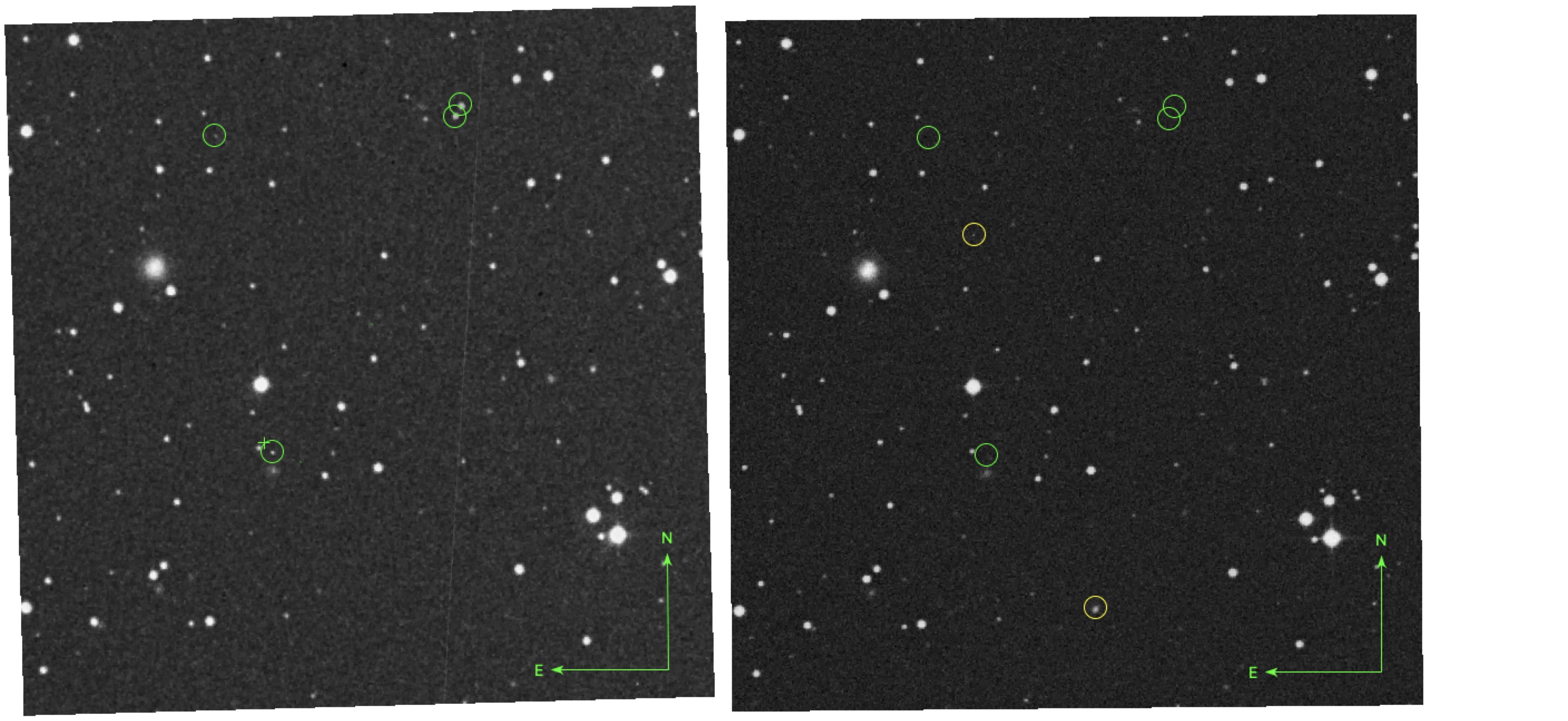}
  \caption{\label{cand1} {\bf Candidate 1.} We show the candidate in SuperCosmos scans of POSS-I red (left) and POSS-II red (right) images. Transients are marked with green circles. The candidate with a measured coordinate is marked with a cross (+). Yellow circles show defects. Also the white line crossing the POSS-I field is a scanning defect. Four transients are visible in the POSS-I image, where three follow a straight line. Box size is 10 x 10 arcmin$^{2}$. See Fig 7 for a version with drawn lines that shows the possible alignment.}
   \end{figure*}
   
   \begin{figure*}
   \includegraphics[scale=0.2]{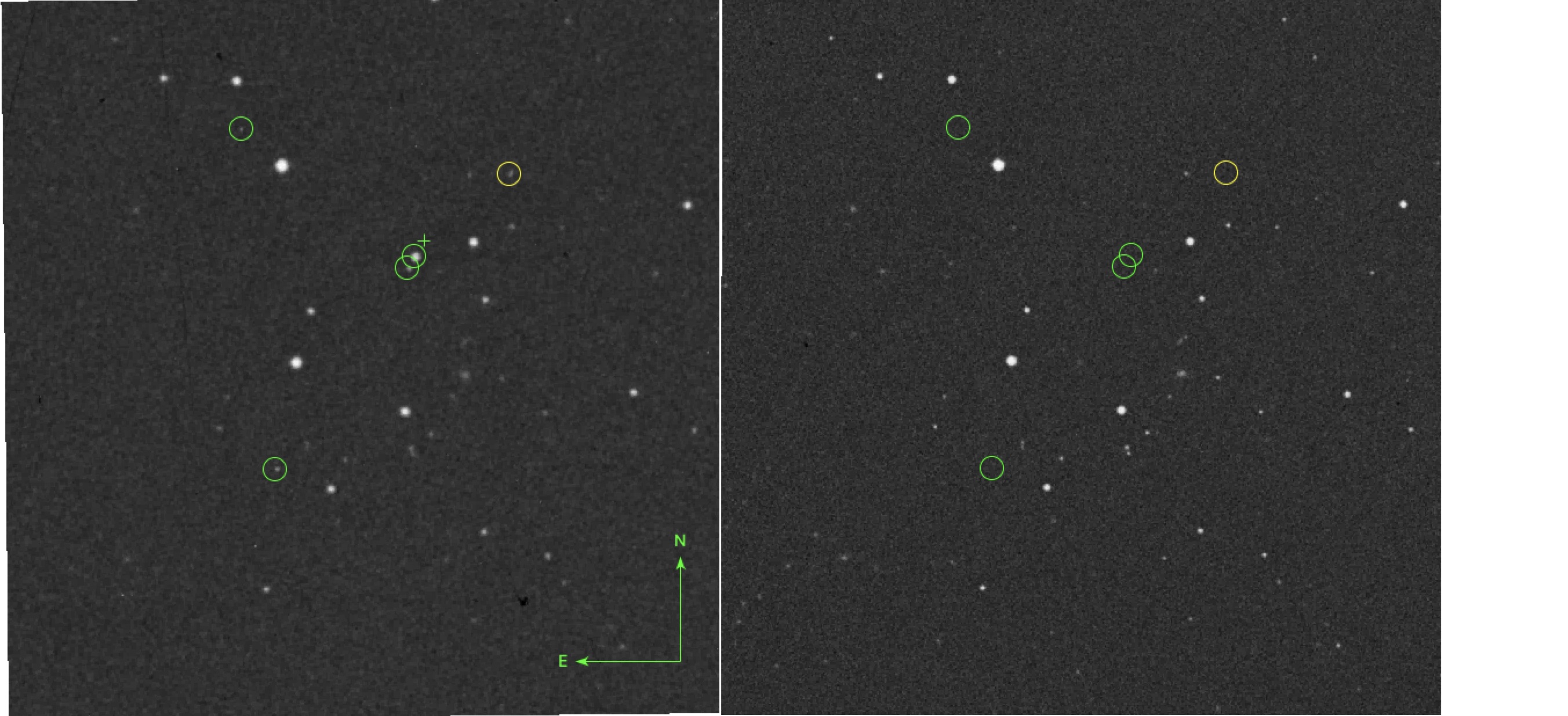}
  \caption{\label{cand2} {\bf Candidate 2.} We show the candidate in SuperCosmos scans of POSS-I red (left) and POSS-II red (right) images. Transients are marked with green circles. The candidate with a measured coordinate is marked with a cross (+). Four transients are visible in the POSS-I image, where three follow a straight line. See Fig 8 for a version with drawn lines that shows the possible alignment. Box size is 10 x 10 arcmin$^{2}$.}
   \end{figure*}
   
    \begin{figure*}
   \includegraphics[scale=0.2]{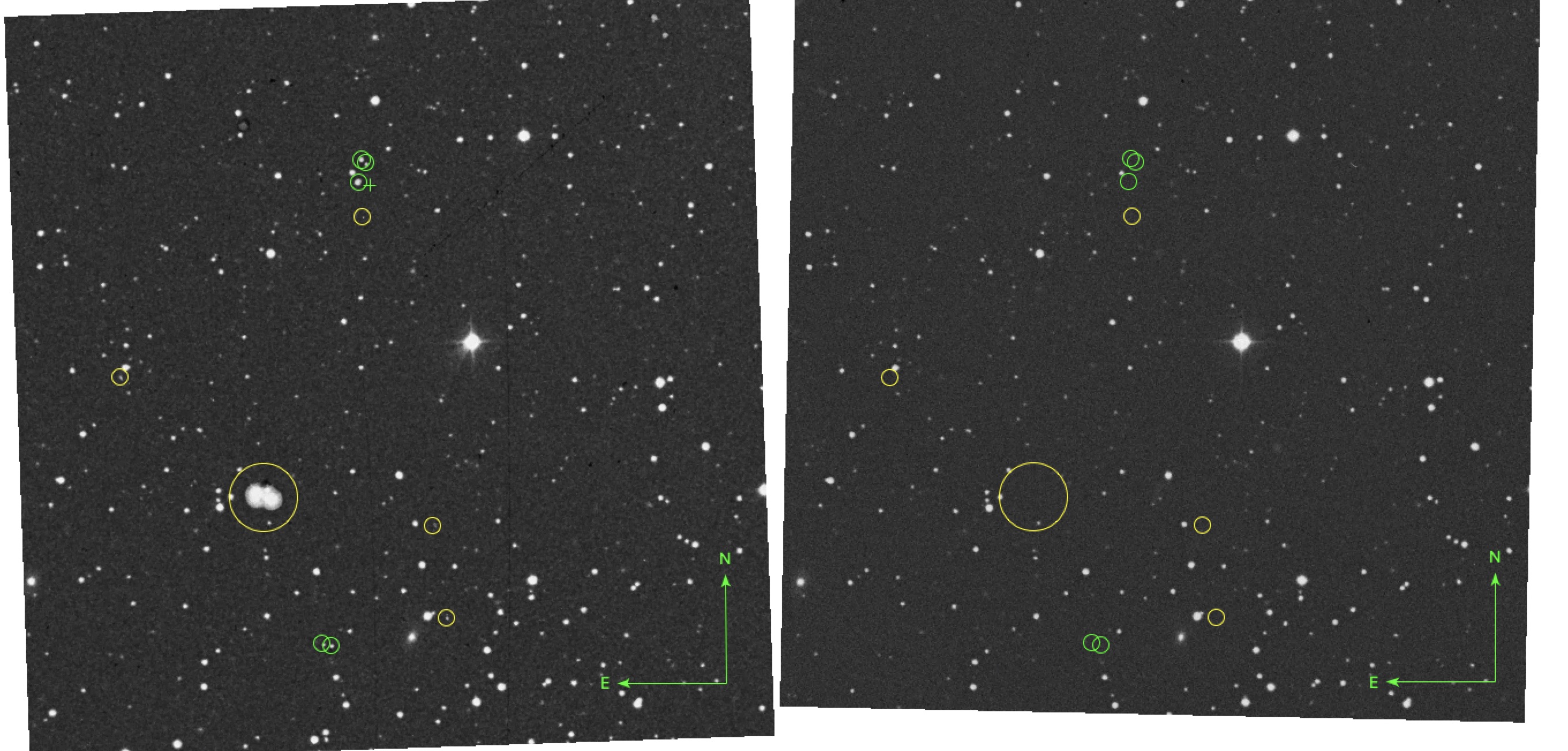}

  \caption{\label{cand3} {\bf Candidate 3.} We show the candidate in SuperCosmos scans of POSS-I red (left) and POSS-II red (right) images. Transients are marked with green circles. The candidate with a measured coordinate is marked with a cross (+) and might be slightly dubious in shape. Yellow circles show defects, both plate defects and scanning defects. See Fig 9 for a version with drawn lines  that shows the possible alignment. Box size is roughly 15 x 15 arcmin$^{2}$.}
   \end{figure*}
   
    \begin{figure*}
   \includegraphics[scale=0.18]{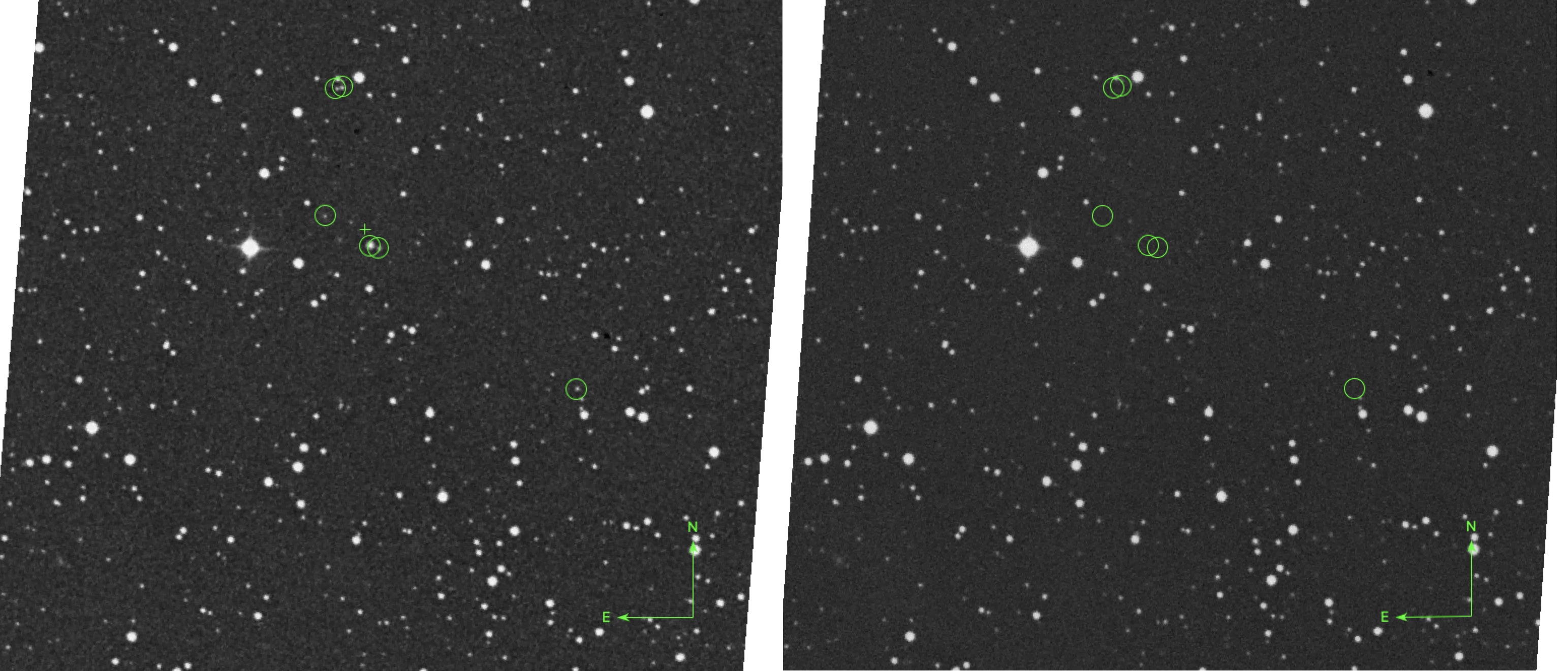}

  \caption{\label{cand4} {\bf Candidate 4.} We show the candidate in SuperCosmos scans of POSS-I red (left) and POSS-II red (right) images. Transients are marked with green circles. The candidate with a measured coordinate is marked with a cross (+). See Fig 10 for a version with drawn lines that shows the possible alignment. Box size is 12 x 12 arcmin$^{2}$.}
   \end{figure*}
   
       \begin{figure*}
   \includegraphics[scale=0.2]{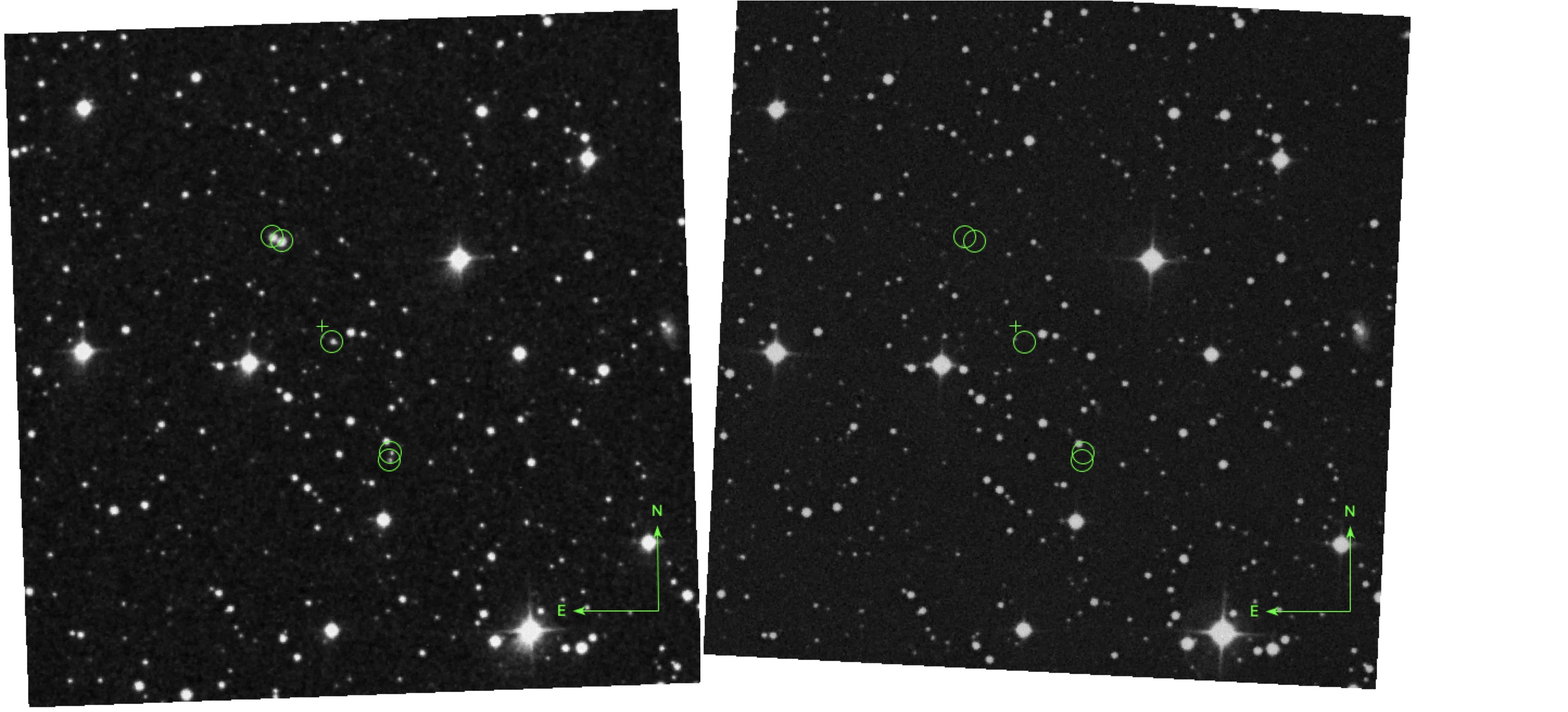}

  \caption{\label{cand5} {\bf Candidate 5.} We show the candidate in SuperCosmos scans of POSS-I red (left) and POSS-II red (right) images. Transients are marked with green circles. The candidate with a measured coordinate is marked with a cross (+). See Fig 11 for a version with drawn lines that shows the possible alignment. Box size is 10 x 10 arcmin$^{2}$.}
   \end{figure*}

\section{Alternative explanations}\label{sec:alternative}

A summary of excluded or improbable astrophysical, observational or instrumental explanations for multiple simultaneous transients, is given in \cite{Villarroel2021a}. Assuming that these simultaneous transients are real, not plate defects, we here explore some alternative explanations.

Point sources could be produced if sunlight is reflected by a highly reflective surface or if the objects are emitting light. As was previously shown in \cite{Villarroel2021a}, such objects must be inside our Solar System. We consider four options: (i) the objects are inside our atmosphere, (ii) the objects are in Low Earth Orbits, (iii) the objects are in geosynchronous orbits, (iv) the objects are even significantly further away from the Earth.

Moving objects inside our atmosphere that either are reflecting light or emitting light  will leave trails as they move during the typical $45 - 50$ minutes exposure of the POSS-I images, which is inconsistent with the several point sources detected in the image, see \cite{Villarroel2021a}. The only situation where several objects inside our atmosphere could produce multiple transient events, is if light-emitting objects simultaneously appear on the sky, do not move at all, and vanish altogether within the plate exposure time. Objects that are close to the observer could be slightly out of focus in an image. Maybe, these transients are caused by a very rare celestial phenomenon that appears as clusters of light appearing and vanishing. We cannot exclude this extraordinary possibility \citep{Villarroel2022b}. Perhaps red sprites high up in the troposphere or rare and poorly understood atmospheric phenomena similar to Hessdalen's clusters of ball lightning could be possible explanations \citep{Teodorani}, and even explain the dubious shapes of some transients among Candidates 3 and 5.

It is also difficult to explain the phenomenon as events at the Low Earth Orbits. The problem of continuous (or missing) illumination of the objects also applies at Low Earth Orbits. We can therefore exclude the possibility of glints appearing as a result of experimental military rockets and missiles that reached only $100-200$ km above the Earth surface.

We also consider much higher altitudes than those of the geosynchronous orbits. As shown in \cite{Villarroel2021a}, asteroids will either leave trails across the plate if they move fast, or if moving slowly by leave point sources in all images taken close in time. But they cannot leave point sources in one long-exposure image, and just half an hour later be entirely absent (an object moving this fast outside the image frame will leave a streak). Even tumbling, \textit{'Oumuamua}-like objects will leave trails. 

Clearly with our current data we cannot consider all possible scenarios to explain the existence of simultaneously appearing and disappearing point-like objects on the sky, as some explanations may not yet have even been imagined.

\section{Object properties}\label{sec:properties}

In the following sections we discuss the conditions under which reflections from objects at the GEO can be produced.
It is interesting to consider what shapes of objects are capable of creating the observed glints. A very fast spinning object might glint many times during the 50 minutes of exposure time, while an object spinning significantly slower might glint only once or twice.

If the object is fast spinning and the strip length $d_{max}$ corresponds to the path travelled by the object, one can calculate the typical speeds and we get a velocity of the order of $\sim$ 0.5 arcsec/second, significantly slower than the typical value at the GEO ($\sim$ 15 arcsec/second). Under these circumstances, one may expect that more transients could be visible along the same line, if just extending the image and searching for more glints along the same line. However, if the spinning is very slow and the complex geometry contains only a few, small shiny surfaces covering a small area relative to the rest of the structure, the glints might happen only for a fraction of the entire exposure time.

\begin{figure*}
   \includegraphics[scale=0.15]{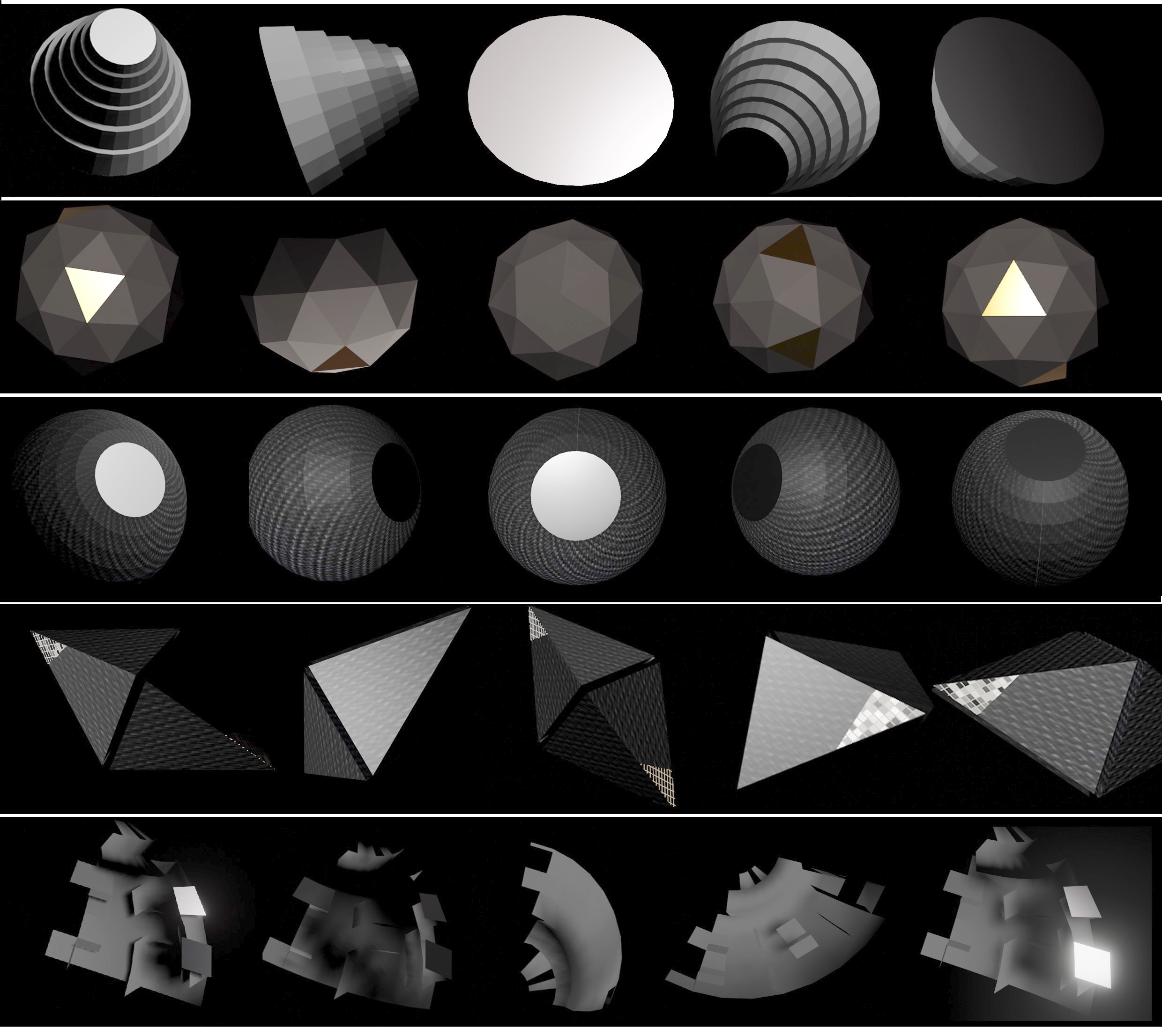}
  \caption{\label{shapes} {\bf Simulated shapes.} We show five different shapes that under slow spinning could produce a handful of glints and in particular double glints. Each shape has two highly reflective surfaces. From top to bottom: (a) cone-like shape, (b) multifaces shape, (c) sphere, (d) 3D hexagone, (e) piece of debris. Each object has both dull and reflective materials on its surface, painted in grey respective light tones. Each object spins around an axis that also has precession, causing the reflective surface not to be visible at all times.}
   \end{figure*}

We simulate several possibilities of shapes and whether they could produce the observed ``glinting patterns'' with help of the \textit{Blender}\footnote{www.blender.org} computer graphics software: a sphere, a three-dimensional multifaced object, a cone, double pyramids and a shape with two-reflective parts. Each shape is made of dull materials combined with a few reflective surfaces, that when they coincide perfectly with the observer and the Sun, give a short but powerful glint. Each object might in addition be spinning on an axis, that may have some precession which makes the glints not visible at all times. We show the five shapes in Figure \ref{shapes}.

It is easy to show that a spherical shape will not produce short glints -- there needs to be some flat reflective surfaces for this. In our simulation we assume the top and bottom surface of the cone are reflective in order to obtain double glints in every rotation. If we add precession to the axis this allows the glints only to appear a few times before the object again has disappeared. 

For the case of the double pyramid, we consider the following scenario: a shiny double pyramid which loses its reflectivity over time until only small areas of it are reflective. Given a spin and precession, only occasionally are these surfaces visible, resulting in a few glints. Thus, each of the five presented surfaces are, in principle, capable of reproducing a glinting pattern similar to the observed one.

\section{The background density}\label{sec:density}

We have identified 83 $r$-point alignments within the northern hemisphere alone. Assuming isotropy we can estimate $\sim$ 170 $r$-point alignments as a lower limit for the whole sky. These alignments have been sampled during a period of 16 years when the POSS-I survey was undertaken between $1949-1958$ \citep[][]{Minkowski1963}. POSS-I sampled the entire northern hemisphere and the southern up to declinations $>-45$ degrees. During this time period 936 ``red'' fields were obtained, for $\sim$ 50 minutes exposure each giving a total if 780 hours exposure time. Each exposure covers $6 \times 6$ sq. degrees each, which means roughly 80\% of the sky has been covered.

If we take the 170 cases to estimate a rough detection rate, we can simply divide 170 by 780 hours and sky coverage, which gives us $\sim 0.27$ hour$^{-1}$ sky$^{-1}$. It is higher than the $\sim 0.07$ events hour$^{-1}$ sky$^{-1}$ estimated in \cite{Villarroel2021a}. It is also significantly lower than the typical glint events $\sim$ 1800 hour$^{-1}$ sky$^{-1}$ \citep{McDowell,Corbett2020} arising from human space debris and satellites observed from the equator, which is why it would be nearly impossible to detect this background population of objects unless it is specifically looked for.

We can also calculate the actual number density of objects.
If we assume that the population of objects has a uniform number per surface unit ($n$) then we know that the number of objects ($N$) detectable at any given time is given by

\begin{equation}
N=n\times S
\end{equation}

where $S$ is the spherical survey surface containing the observed reflective objects:

\begin{equation}
    n  =\frac{N}{S},\\
\end{equation}

where the spherical surface area $S=4\pi d^2$ is calculated for the radius of a geosynchronous orbit $d$, so that:

\begin{equation}  
    n  =\frac{N}{4\pi d^2}.\\
\end{equation}

We set $d=42164$ km as the radius of the geosynchronous orbits. Using $N=170$ for the number of detected $r$-point alignments we find that 
\begin{equation}
n=3.76\times10^{-9}\,{\rm objects \,\,  km}^{-2}
\end{equation}

These estimates provide a guide to the number of background objects that may exist inside the surveyed volume. However, not every object will produce several glints. The shape and the reflectivity of an object will determine the likelihood for one or more glints. This uncertainty also leads to an underestimation of the number density of objects, which could actually be even one order of magnitude higher.

\section{Discussion}\label{sec:discussion}

Does there exist a background population of objects with at least one flat, reflective surface in geosynchronous orbits around Earth? This is the fundamental question addressed by the present study.

We use a simple and direct method to detect these hypothetical objects, by searching for several transients aligned along a line in a dataset devoid of known human-made satellites and space debris. This follows the principle of searches for non-terrestrial artifacts through a \textit{smoking gun indicator} as outlined in \citep{Villarroel2021a}. For practical purposes we start with the list of VASCO transients in the northern hemisphere provided by \cite{Solano2021}. We find $\sim$83 initial candidate $r$-point alignments in the northern hemisphere. We also find several systems with double or triple transients. The triplets are interesting, as they are consistent with reflections off flat, rotating surfaces \citep{Deil2009}.

We carefully investigate each of the 22 cases of potential 4- and 5-point alignments (some of which are reduced to 3-point alignments upon further inspection) and list the five most interesting cases based upon the level of statistical significance of each alignment (see Section \ref{sec:statistics}). Nevertheless, the uncertainties are too big to estimate a \textit{total} probability to find a particular alignment in our study (needed in order to establish independently if an event is a GEO glint). All what we can extract is the probability that a particular candidate alignment is non-random in a single image. For these cases, the choice of FWHM matters and in best case the probabilities for the five candidate alignments range between $2.5$ and $4 \sigma$. The best candidates turn out to be three 4- and 5-point alignments, of which two have dubious shapes on their transients.
We cannot exclude that among the 61 3-point alignments some might be of interest in case they are proven authentic. Overall, based on statistics alone the results are inconclusive. In order to establish whether or not these examples are authentic, the shortlisted candidates on their original photographic plates must be studied under a microscope.

Within the theoretical framework of our experiment, we use the assumed glinting patterns to infer a possible shape, shown in Figure \ref{shapes}. A combination of a multifaceted structure with a few reflective pieces together with slow rotation around an axis together with precession, can reproduce the observed glinting patterns. We further infer a detection rate of $\sim0.27$ events hour$^{-1}$ sky$^{-1}$ for aligned multiple transients. The corresponding surface density of detectable objects is $3.8\times10^{-9}$ objects km$^{-2}$, where the actual density of objects could be even an order of magnitude higher. 

Even if these events were real observations of simultaneous flashes on the sky (rather than any contamination of the photographic plates), their association with objects in geosynchronous orbits glinting, might not be correct. The strongest objection to the GEO hypothesis is that we have not yet found evidence for any event with a periodicity or quasi-periodicity, in the glinting pattern. Naturally, an object spinning very slowly will only leave a single or a small number of glints in an image and such objects might be slow spinners. Whether or not an object has a periodic glinting pattern depends on the complexity of its shape and surface and if it has more than one rotation axis. We have not examined whether there might be more aligned transients found by extending the stripe far beyond the image frame as this is a considerable challenge, and complicates the statistical analysis.

In addition, the transients may be elongated if they are, in fact, moving across the sky, as the GEO hypothesis states. Transients might also be elongated if the objects are larger than some tens of meters or if they are further than some $\sim$ 100 000 km from the Earth. In the future we will extend our searches to ``dashed lines'' in the POSS-I dataset. A less perceivable elongation for apparent ``stars'' could also be found using high-resolution digitizations of the relevant plates, the axis of elongation can be measured and its direction can be compared with the line of alignment between the transients. In the present work, we have specifically looked for non-elongated transients.

Summarizing, from the present analysis we have only a few examples, that all need to be subjected to a careful examination under a microscope in order to establish or reject their authenticity. But even if simultaneous transients are real, the eventual explanation may be something which is currently inconceivable. Some alternative explanations are presented in Section \ref{sec:alternative}.

\section{Moving forward}

There are several ways forward. The first is to investigate each of the $\sim$ 5 candidate alignments and other multiple transient events under the microscope, using the original photographic plates stored at Caltech. If a single of the multiple transient-candidates survives the examination under a microscope, we can confirm multiple transients; a single verified case of multiple transients, aligned or unaligned, is all that is needed to confirm the existence of the phenomenon of multiple transients. Under the circumstance that every reported instance of multiple transients instead is \textit{rejected}, the estimated surface density of objects is directly transformed into an upper limit of $<10^{-9}$ km$^{-2}$ non-terrestrial artifacts in orbit around Earth. This limit is valuable for space archeology programs such as the \textit{Galileo} project\footnote{https://projects.iq.harvard.edu/galileo} that aim to search for non-terrestrial artifacts in the close presence of our planet (even inside the atmosphere). The disadvantage with this process is that the Caltech plates are difficult to access which makes the experiment more difficult to reproduce.

Furthermore, we wish to examine other photographic plate surveys. Repeating the same experiment with a subset of the Harvard DASCH plates \citep{Grindlay2012}, digitized plates from the Lick observatory, \textit{Carte du Ciel}, Tautenburg observatory \citep{Andruk2021} or the FON Dushanbe catalogue \citep{Kokhirova2021} is fruitful in searches for a rare alignment displaying a periodic or quasiperiodic glinting pattern with larger \textit{r} number. In future studies, we will also relax the search condition in our automated searches, to increase the starting sample of $r$-point alignments. These $r$-point alignments could be visually explored in the citizen science project. All these efforts could help us to either obtain stronger candidates in search for a 5$\sigma$ event to confirm the hypothesis, and obtain more exact estimates on the shapes and possible number of objects involved, if they exist.

The third and most rigorous approach to test the hypothesis, is through the search for such events in current and future digital surveys like the \textit{Vera C. Rubin} observatory that will have a cadence greater than any other survey. Modern astronomical surveys use CCD imaging with seconds-to-minute long exposures. For example, an automated survey like the \textit{Zwicky Transient Facility} (ZTF) scans each segment of the sky once per day. Given that any high-albedo objects are moving, spinning and emitting reflections from time to time, we should be able to confirm their existence through customized searches of modern data. The recently initiated \textit{Galileo} project, has suitable instrumentation and software development needed to search for these glinting events more effectively than was possible with other surveys.
A survey that involves exposures of seconds-to-minutes duration, will enable us to better constrain the limit on the simultaneity of appearance and movement of the glinting pattern of a potential object. This would also provide a more accurate constraint on geometries, properties and the total number density of objects.

\section{Conclusions}

This paper summarizes a first set of searches for multiple, simultaneously (within a plate exposure) appearing and vanishing optical point sources in the night sky with an internal alignment. A few candidates are presented in the paper, but to confirm their authenticity the original photographic plates must be examined under a microscope to exclude point-like plate defects causing the ``transients''.

We speculate on the properties of a background population of objects with flat, highly reflective surfaces in geosynchronous orbits around the Earth. The expected detection rate of aligned transients from such a population corresponds to $\sim 0.27$ hour$^{-1}$ sky$^{-1}$. The surface density of the objects that range in size from cm to a few tens of meters corresponds to $3.8\times10^{-9}$ objects km$^{-2}$.
But, if the transients all turn out to be false positives on a photographic plate, this means we have probed the Earth's nearest vicinity (from a few thousands km to $100,000$ km) for non-terrestrial artifacts and found not a single object at high orbits around the Earth, yielding a upper limit of $< 10^{-9}$ objects km$^{-2}$ near our planet.

Finally, there are still uncertainties that preclude a definite answer even if the authenticity of the transients eventually is confirmed, and one of these is to fully understand the phenomenology behind simultaneously appearing and disappearing point sources, that may originate in a entirely different process. Thus, the firm notion of a background population of high-albedo objects in geosynchronous orbits around Earth must await empirical confirmation by means of further studies of both the historical data, and comparisons with the new deep electronic surveys of variable phenomena.

\section{Acknowledgments}\label{sec:ackn}

B.V. wishes to first and foremost thank Geoffrey W. Marcy for many inspiring discussions and great suggestions on the draft. It is much appreciated. B.V. also wishes to thank Avi Loeb (Galileo project), Robert Powell (SCU/Galileo), Sarah Little (SCU/Galileo) and Vitaly Andruk for helpful and constructive comments.

The Digitized Sky Surveys were produced at the Space Telescope Science Institute under U.S. Government grant NAG W-2166. The images of these surveys are based on photographic data obtained using the Oschin Schmidt Telescope on Palomar Mountain and the UK Schmidt Telescope. The plates were processed into the present compressed digital form with the permission of these institutions.
The National Geographic Society - Palomar Observatory Sky Atlas (POSS-I) was made by the California Institute of Technology with grants from the National Geographic Society.
The Second Palomar Observatory Sky Survey (POSS-II) was made by the California Institute of Technology with funds from the National Science Foundation, the National Geographic Society, the Sloan Foundation, the Samuel Oschin Foundation, and the Eastman Kodak Corporation. The Oschin Schmidt Telescope is operated by the California Institute of Technology and Palomar Observatory. The UK Schmidt Telescope was operated by the Royal Observatory Edinburgh, with funding from the UK Science and Engineering Research Council (later the UK Particle Physics and Astronomy Research Council), until 1988 June, and thereafter by the Anglo-Australian Observatory. The blue plates of the southern Sky Atlas and its Equatorial Extension (together known as the SERC-J), as well as the Equatorial Red (ER), and the Second Epoch [red] Survey (SES) were all taken with the UK Schmidt. All data are subject to the copyright given in the copyright summary. Copyright information specific to individual plates is provided in the downloaded FITS headers. Supplemental funding for sky-survey work at the STScI is provided by the European Southern Observatory.

This research has made use of the Spanish Virtual Observatory (http://svo.cab.inta-csic.es) supported from Ministerio de Ciencia e Innovación through grant PID2020-112949GB-I00. B.V. is funded by the Swedish Research Council (Vetenskapsr\aa det, grant no. 2017-06372) and is also supported by the The L’Or\'{e}al - UNESCO For Women in Science Sweden Prize with support of the Young Academy of Sweden. She is also supported by M\"{a}rta och Erik Holmbergs donation. Nordita is partially supported by Nordforsk. M.E.S. acknowledges financial support from the Annie Jump Cannon Fellowship, supported by the University of Delaware and endowed by the Mount Cuba Astronomical Observatory.

   \begin{figure*}
   \includegraphics[scale=0.2]{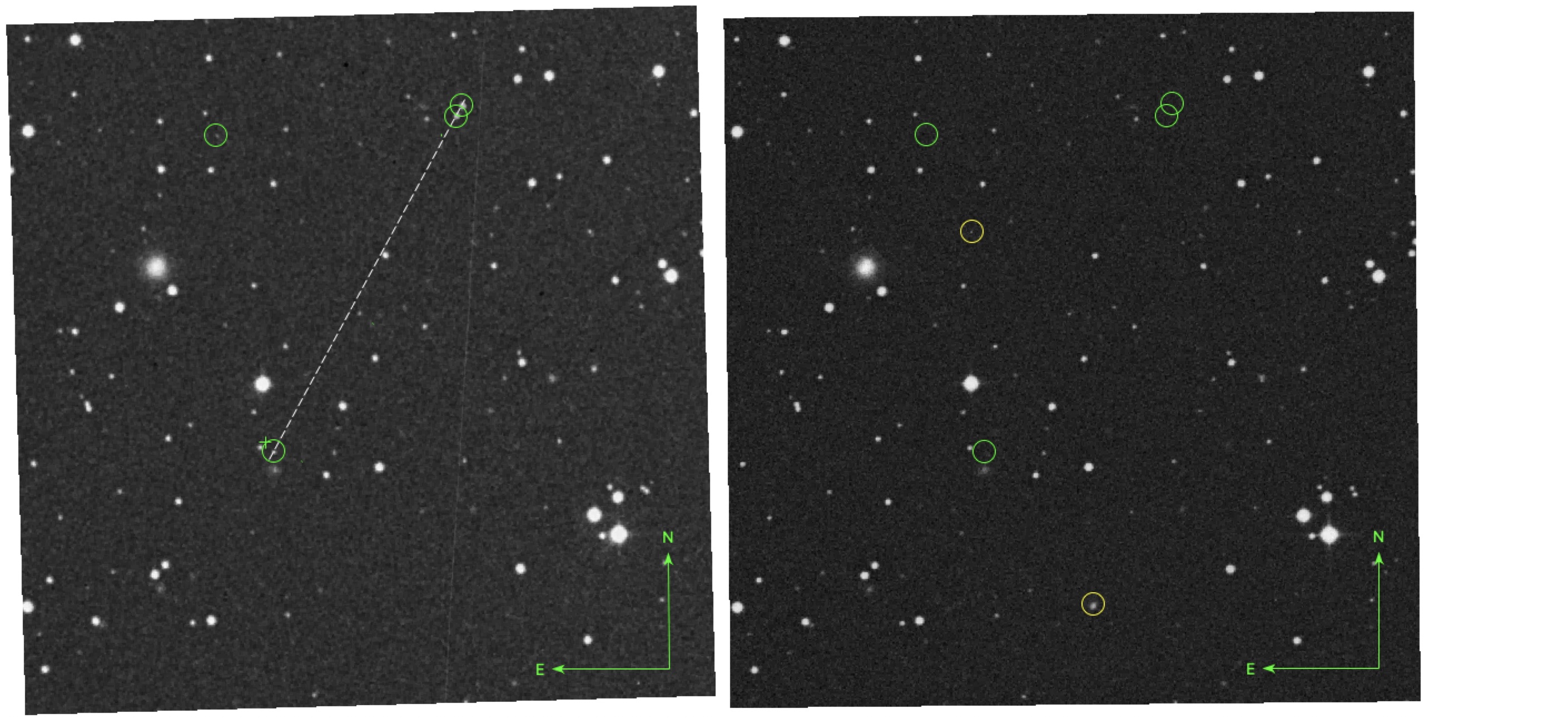}
  \caption{\label{psf1} {\bf Candidate 1.} We show the candidate in SuperCosmos scans of POSS-I red (left) and POSS-II red (right) images. Transients are marked with green circles. The candidate with a measured coordinate is marked with a cross (+). A dashed white line shows the alignment. Yellow circles show defects. Also the white line crossing the POSS-I field is a scanning defect. We see 4 transients in the POSS-I images where three follow a straight line.}
   \end{figure*}
   
   \begin{figure*}
   \includegraphics[scale=0.2]{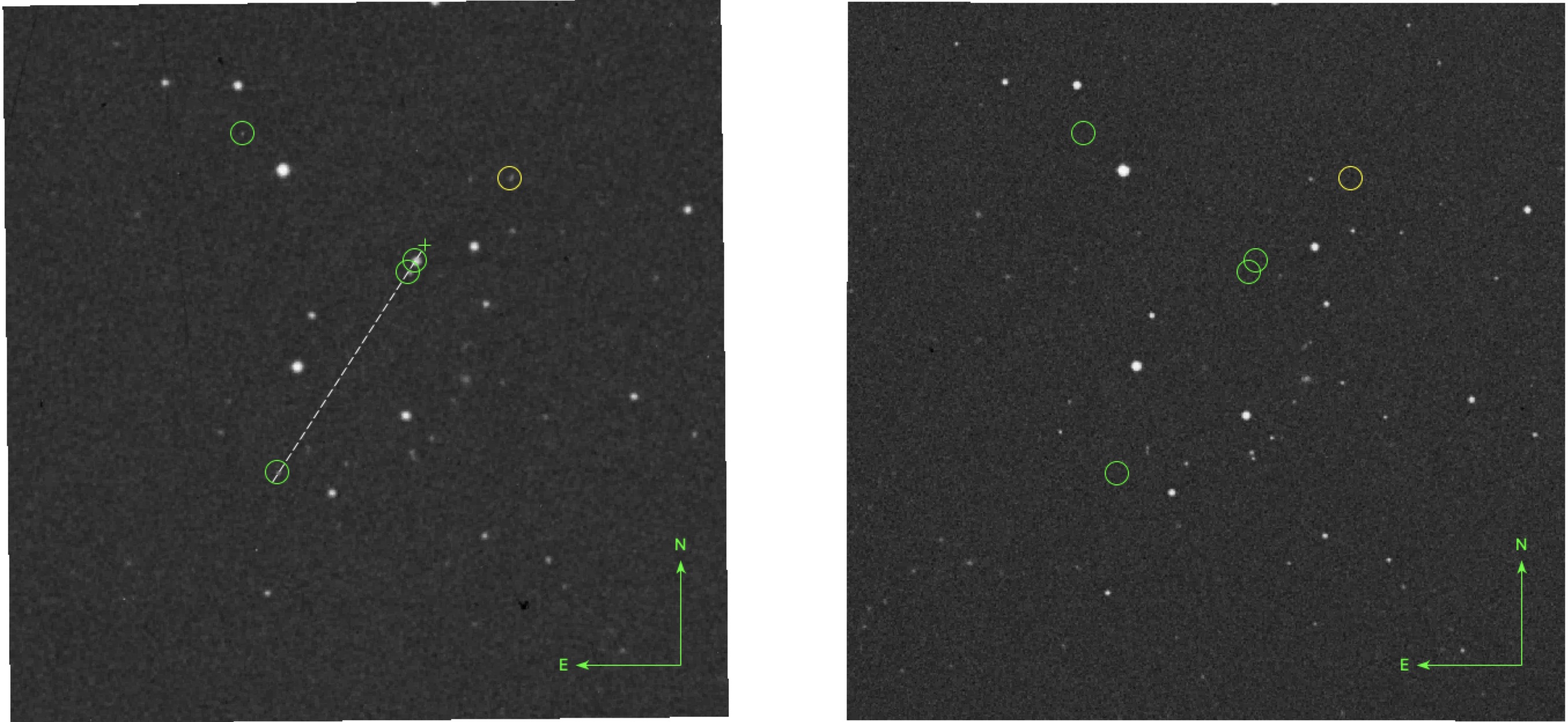}
  \caption{\label{psf2} {\bf Candidate 2.} We show the candidate in SuperCosmos scans of POSS-I red (left) and POSS-II red (right) images. Transients are marked with green circles. The candidate with a measured coordinate is marked with a cross (+). A dashed white line shows the alignment. Yellow circles show defects. Also the white line crossing the POSS-I field is a scanning defect. We see 4 transients in the POSS-I images where three follow a straight line.}
   \end{figure*}
   
    \begin{figure*}
   \includegraphics[scale=0.2]{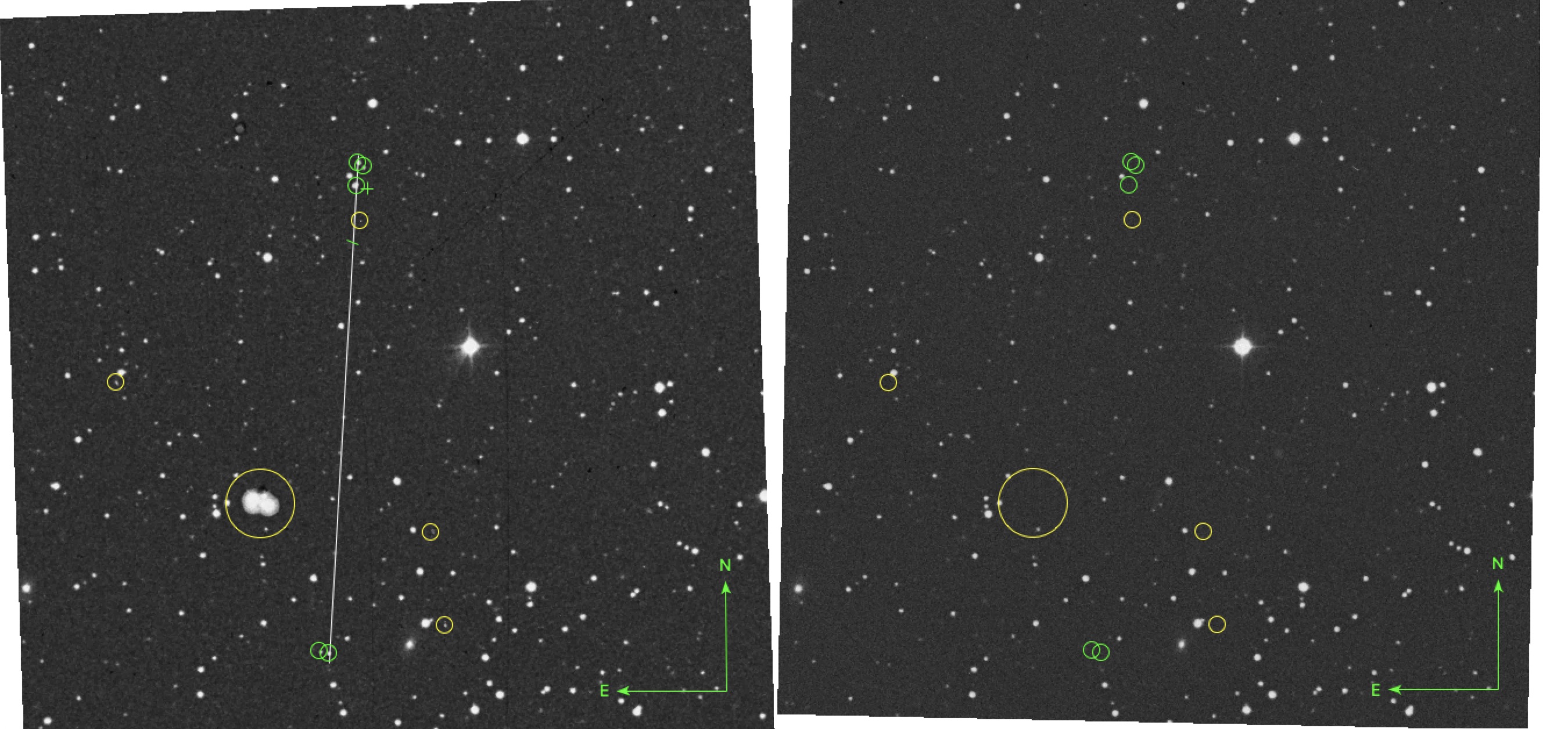}
\includegraphics[scale=0.2]{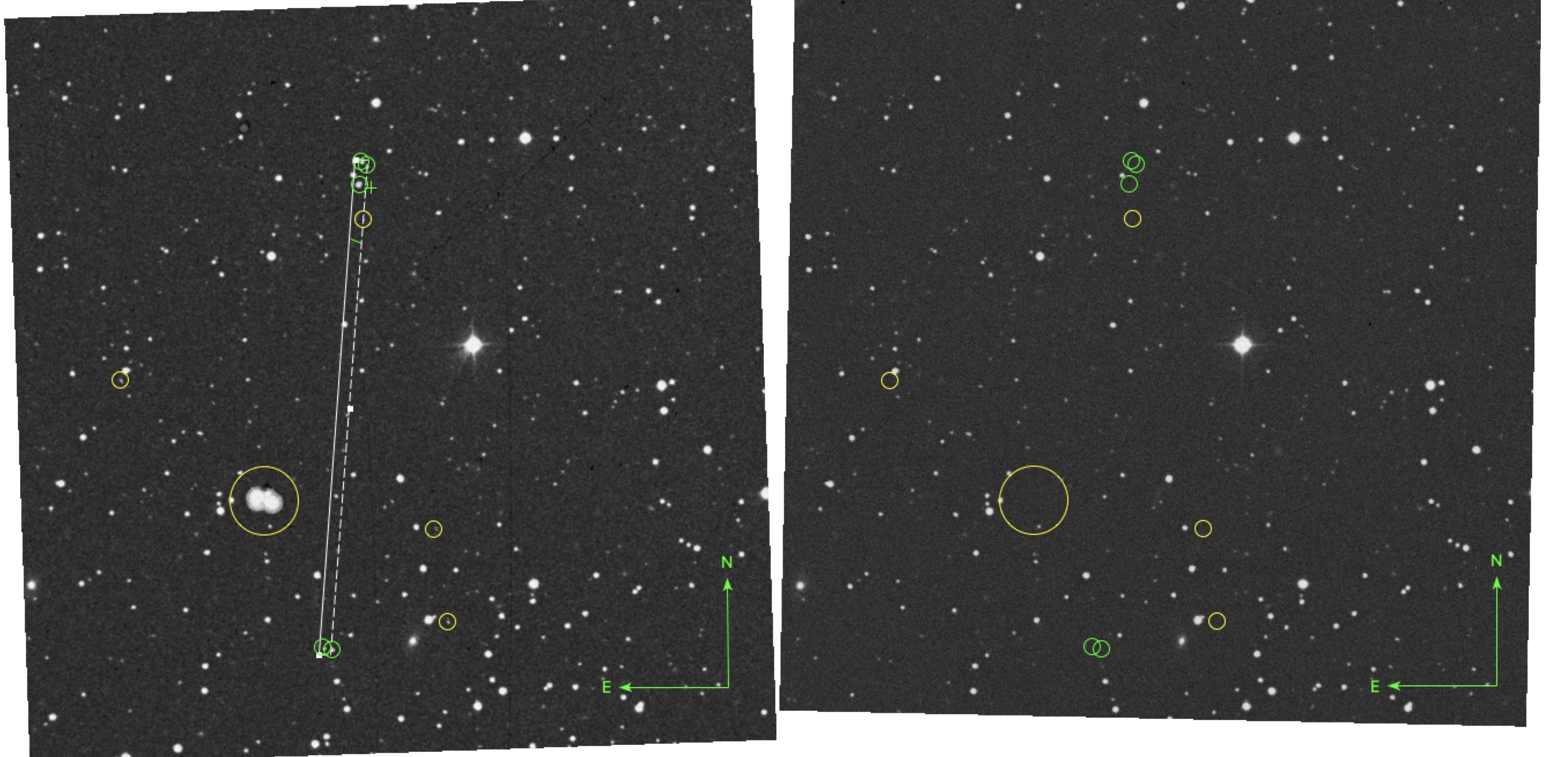}

  \caption{\label{psf3} {\bf Candidate 3.} We show the candidate in SuperCosmos scans of POSS-I red (left) and POSS-II red (right) images. The upper row shows a 3-point alignment within 1 - 2 arcsec. The lower row shows a 5-point alignment of within 15 arcsec. Transients are marked with green circles. The candidate with a measured coordinate is marked with a cross (+) and might be slightly dubious in shape. The dashed lines shows the alignment (the white double line for the thicker alignment below). Yellow circles show defects, both plate defects and scanning defects.}
   \end{figure*}
   
    \begin{figure*}
   \includegraphics[scale=0.18]{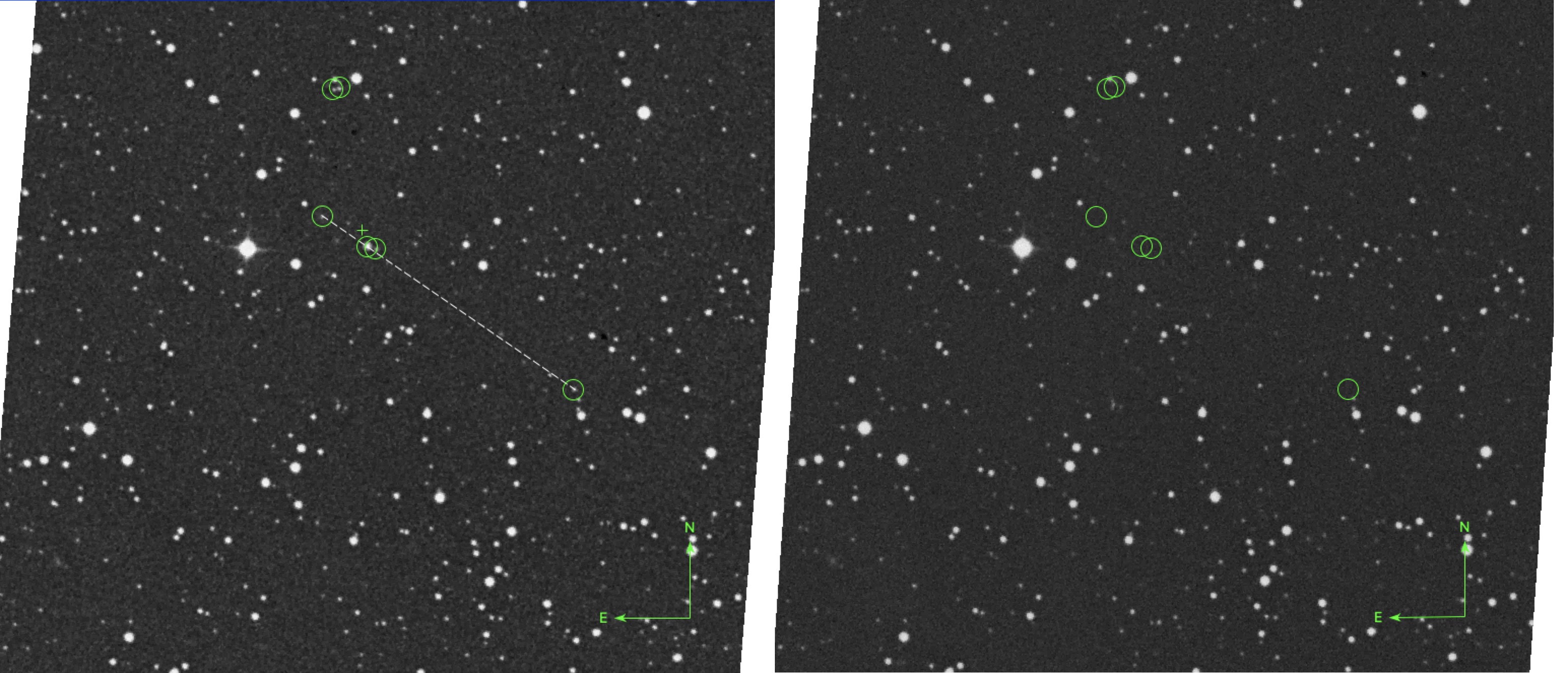}
\includegraphics[scale=0.18]{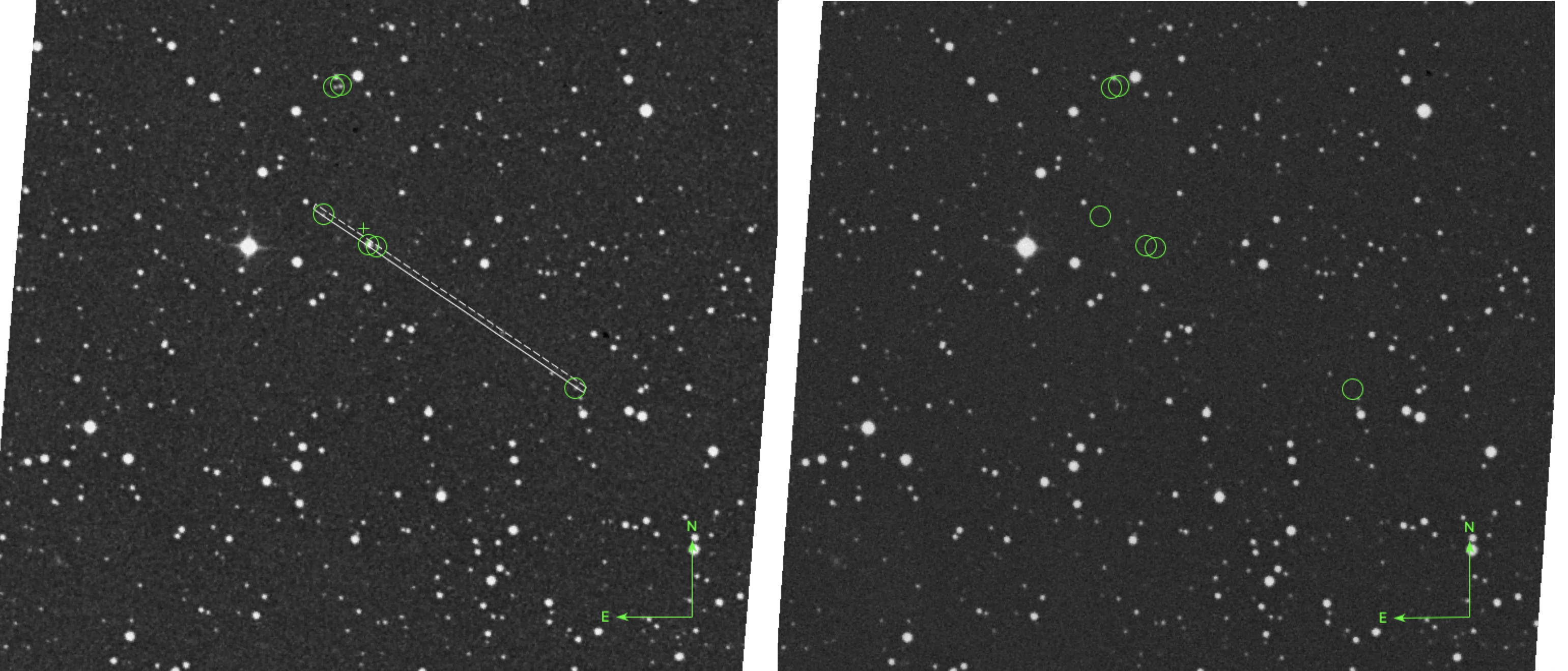}

  \caption{\label{psf4} {\bf Candidate 4.} We show the candidate in SuperCosmos scans of POSS-I red (left) and POSS-II red (right) images. The upper row shows a 3-point alignment within 1 arcsec. The lower row shows a 4-point alignment of within 5 arcsec. Transients are marked with green circles. The candidate with a measured coordinate is marked with a cross (+). The dashed lines shows the alignment (the white double line for the thicker alignment below).}
   \end{figure*}
   
       \begin{figure*}
   \includegraphics[scale=0.2]{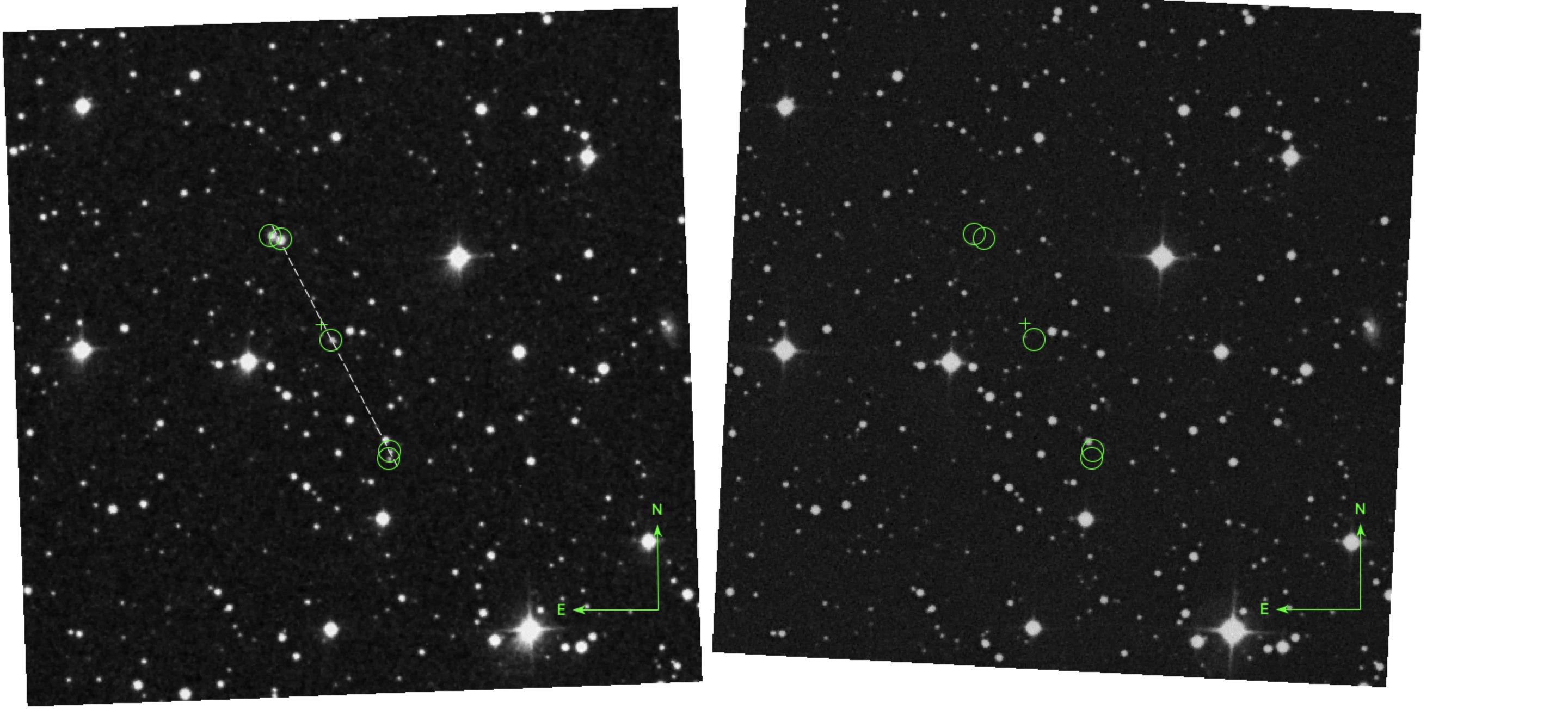}
\includegraphics[scale=0.2]{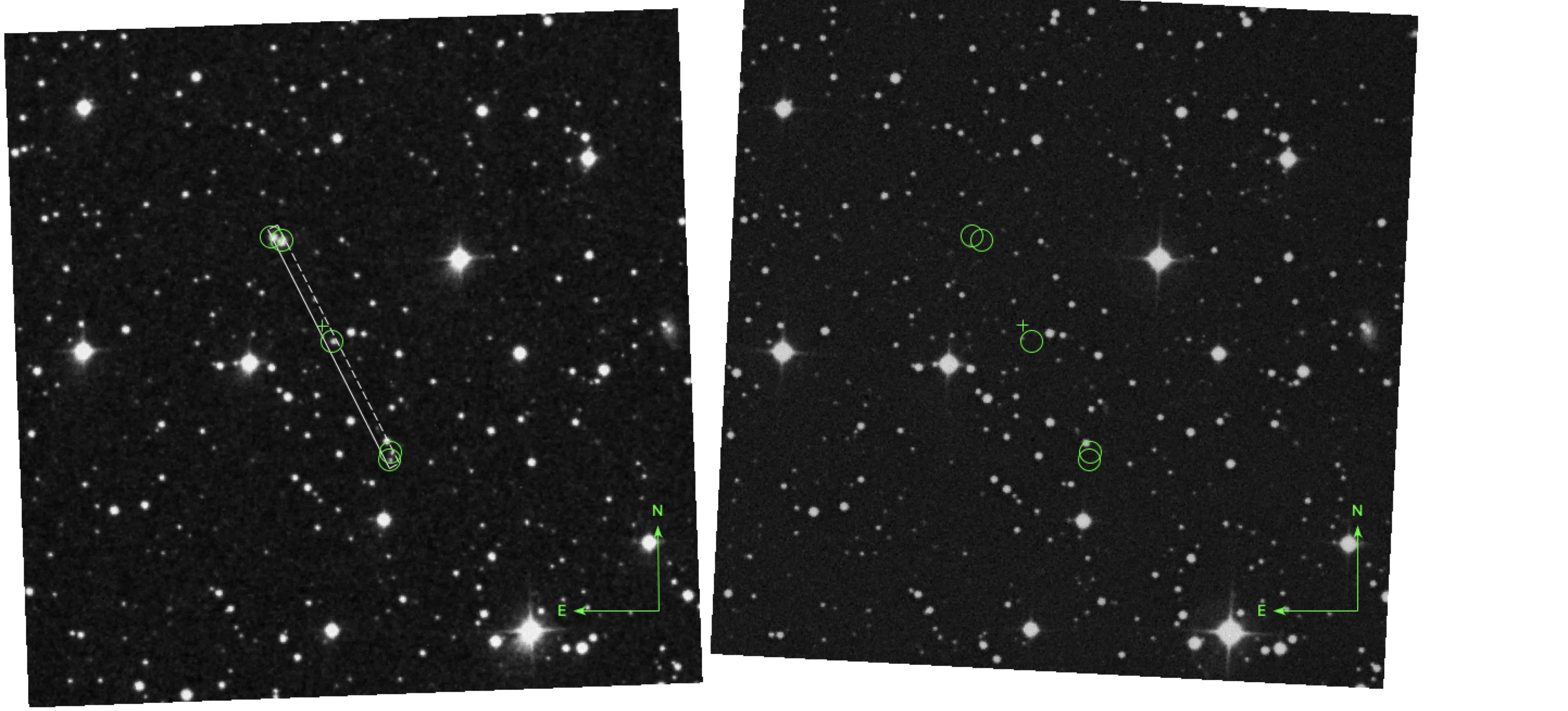}

  \caption{\label{psf5} {\bf Candidate 5.} We show the candidate in SuperCosmos scans of POSS-I red (left) and POSS-II red (right) images. The upper row shows a 3-point alignment within 1 arcsec. The lower row shows a 5-point alignment of within 10 arcsec. Transients are marked with green circles. The candidate with a measured coordinate is marked with a cross (+). The dashed lines shows the alignment (the white double line for the thicker alignment below).}
   \end{figure*}





\end{document}